\documentclass[reprint, amsmath,amssymb, aps,
]{revtex4-2}

\usepackage[utf8]{inputenc}
\usepackage{ragged2e}
\usepackage{array}
\usepackage{comment}
\usepackage{graphicx}
\usepackage{multirow}
\usepackage{amsmath}
\usepackage{amssymb}
\usepackage[none]{hyphenat}
\usepackage{xcolor}
\usepackage{hyperref}  
\usepackage{natbib}
\usepackage{tcolorbox}
\usepackage{enumitem}
\usepackage{tikz}
\usepackage{tikz-3dplot}
\usetikzlibrary{positioning}
\hypersetup{colorlinks=true,urlcolor=blue,citecolor=blue,linkcolor=blue}
\usepackage[normalem]{ulem}

\begin{document}
\sloppy
\preprint{APS/123-QED}

\title{Feedback That Clicks: Introductory Physics Students' Valued Features in AI Feedback Generated From Self-Crafted and Engineered Prompts}

\author{Amogh Sirnoorkar}
\affiliation{Department of Physics and Astronomy, and \\ Department of Curriculum and Instruction, Purdue University, West Lafayette, Indiana - 47907}

\author{N. Sanjay Rebello}
\affiliation{Department of Physics and Astronomy, and \\ Department of Curriculum and Instruction, Purdue University, West Lafayette, Indiana - 47907}


\begin{abstract}

Since the advent of GPT-3.5 in 2022, Generative Artificial Intelligence (AI) has shown tremendous potential in STEM education, particularly in providing real-time, customized feedback to students in large-enrollment courses. A crucial skill that mediates effective use of AI is the systematic structuring of natural language instructions to AI models, commonly referred to as prompt engineering. This study has three objectives: (i) to investigate the sophistication of student-generated prompts when seeking feedback from AI on their arguments, (ii) to examine the features that students value in AI-generated feedback, and (iii) to analyze trends in student preferences for feedback generated from self-crafted prompts versus prompts incorporating prompt engineering techniques and principles of effective feedback. Results indicate that student-generated prompts typically reflect only a subset of foundational prompt engineering techniques. Despite this lack of sophistication, such as incomplete descriptions of task context, AI responses demonstrated contextual intuitiveness by accurately inferring context from the overall content of the prompt. We also identified 12 distinct features that students attribute the usefulness of AI-generated feedback, spanning four broader themes: Evaluation, Content, Presentation, and Depth. Finally, results show that students overwhelmingly prefer feedback generated from structured prompts, particularly those combining prompt engineering techniques with principles of effective feedback. Implications of these results such as integrating the principles of effective feedback in design and delivery of feedback through AI systems, and incorporating prompt engineering in introductory physics courses are discussed.     

\end{abstract}

\maketitle



\section{Introduction}\label{sec:intro}

Providing customized student-facing feedback in large enrollment courses has been one of the emerging topical areas in physics education research~\cite{sirnoorkar2025self}. Feedback is any information provided by an agent (such as people or material artifacts) about one’s performance~\cite{hattie2007power}. The sources of feedback can be external as well as internal. In the context of student learning, information from peers, teachers, algorithms, etc. can act as external sources whereas one's lived experiences, self-monitored observations, etc. can act as internal sources of feedback. This information can take various forms such as corrective, directive, alternative, or affective~\cite{burke2010giving}. Effective  feedback plays a crucial role in improving students’ learning along with enhancing their skills and academic performance~\cite{azevedo1995meta}. Effective feedback is also instrumental in motivating students to actively engage with content knowledge, with qualitative feedback considered more effective than quantitative feedback~\cite{giamos2024continuous}. Researchers thus have called for students' active involvement in the feedback process by facilitating them in interpreting, internalizing, and taking actions based on the provided information~\cite{nicol2006formative}.    

Facilitating effective and engaging external feedback poses a major challenge for educators. Resource demands such as human expertise, material costs, and time constraints make the process challenging. Additional barriers such as the need for timely delivery of information and the incorporation of design features that motivate students to meaningfully engage with and follow up on the information also compound the challenges. Consequently, numerous studies have explored ways to reliably automate the design and delivery of customized feedback based on student responses~\cite{sirnoorkar2023theoretical,laverty2022new,allen2025students}. 

Recent advancements in Generative Artificial Intelligence (henceforth referred as `AI') promise a huge potential in addressing these barriers. This promise stems from AI’s ability to process multi-modal data and generate real time, coherent, customized, responses based on user inputs~\cite{polverini2024understanding,polverini2024evaluating}. The user inputs are natural language instructions, referred to as `prompts', that guide the behavior and output of AI models. With carefully crafted prompts significantly enhancing the AI output, `prompt engineering' has evolved into an empirical and growing body of research focused on ways of systematically structuring prompts to better interact with AI systems. 

Broadly, prompt engineering techniques can be categorized into two categories: foundational and advanced~\cite{chen2025unleashing}. Foundational techniques emphasize clarity, structure, and organization of prompts to optimize outputs. Advanced techniques involve more complex interactions with dynamically responding to AI outputs to yield precise, accurate, and high-quality results (refer Section~\ref{subsec:prompt-engineering} for details). In physics education, researchers have explored the potential of prompt engineering in enhancing accuracy of AI responses to physics assessments~\cite{polverini2024understanding} and prompting approaches of elementary pre-service teachers while seeking appropriate representational formats of physics concepts~\cite{hamed2025dual}. Despite concerns on content accuracy and students' bias towards language sophistication, evolving models have significantly improved in responding to introductory physics content often times even with no emphasis on prompting~\cite{kortemeyer2025boiling,kortemeyer2023could,sirnoorkar2024student}.

The process of feedback generation broadly involves an initial assessment or grading of student responses (typically to an assessment) based on a predetermined rubric, followed by communication of results through performance reports. In physics education, considerable attention has been devoted to the former, i.e., AI-assisted grading of student responses. Studies in this area have leveraged AI models to grade students’ handwritten solutions~\cite{kortemeyer2024grading}, written descriptions of strategies to conceptual questions~\cite{kortemeyer2024performance,kortemeyer2025assessing,lee2024applying,chen2025grading}, and solutions to numerical problems~\cite{wei2025research}. This strand of research broadly notes that while AI models are becoming increasingly effective at grading, their accuracy varies depending on the combination of the employed prompting strategies. Moreover, they also highlight the continued need for active human oversight, particularly when grading problems that involve complex reasoning processes, hand-drawn representations, or responses at the lower end of the score spectrum.  

Research on AI-generated feedback reports in physics education is gradually evolving with studies increasingly focusing on leveraging prompting techniques to design customized, student-facing feedback. Instructors and students alike have appreciated the feedback content~\cite{chen2025grading} both in terms of correctness and usefulness~\cite{wan2024exploring}. Dai {\em et al}.~\cite{dai2025students} observed that mandated versus on-demand use of AI feedback by high school students had varied influences on their academic performance. Additional investigations have thus unpacked students’ perspectives, where students have described AI-generated feedback as useful, actionable, and clearly written, with particular appreciation for suggested improvements and rewritten examples~\cite{mills2025prompting,allen2025students}. 

The above literature on design and delivery of feedback through AI systems highlight several gaps. Firstly, compared to the literature on AI grading (and to benchmarking the performance of AI models on physics assessments), research on AI-generated feedback in physics education remains relatively underexplored. Secondly, studies that focus on AI feedback have primarily employed empirical approaches in designing feedback with little to no focus on embedding the principles of effective feedback established in the education research literature.  Thirdly, research on features of AI-generated feedback perceived as valuable by students while reasoning about physics problems remains relatively unexplored. Lastly, despite the reinforcing results on the significance of prompting approaches on the performance of AI models, coupled with students' increased use of AI, there is a need for investigating the extent of sophistication in student-generated AI prompts when interacting about course-related content.

In the current study, we address the aforementioned gaps in the literature by investigating the following research questions:

\begin{enumerate}[]
\item[\bf RQ 1:] {\em To what extent do students' self-crafted prompts to AI evidence the features of foundational prompt engineering techniques? }

\item[\bf RQ 2:] {\em What features of AI-generated feedback do students perceive as useful when the feedback is based on (a) self-crafted prompts, (b) prompts incorporating foundational prompt-engineering techniques, and (c) prompts combining foundational prompt-engineering techniques with principles of effective feedback?}

\item[\bf RQ 3:] {\em What are the broad trends in students’ preferences for AI-generated feedback based on (a) their self-crafted prompts, (b) prompts entailing foundational prompt-engineering techniques, and (c) prompts combining the foundational prompt-engineering techniques with principles of effective feedback?}
\end{enumerate}

The current article is structured as follows: in the next section, we provide the background on prompt engineering and principles of effective feedback. In Section~\ref{sec:methods}, we describe the study’s data, context, and analysis. In the following section, we present the results before discussing them Section~\ref{sec:discussion}. We then conclude with study's implications, limitations, and directions for future research in Section~\ref{sec:conclusions}.

\section{Background}\label{sec:background}

\subsection{Prompt engineering}\label{subsec:prompt-engineering}

Since the advent of GPT-3.5 in 2022, AI has garnered huge traction among students, teachers, administrators, and researchers alike. This rise stems from the ease of generating diverse and sophisticated multimedia content using simple, natural language instructions, referred to as `prompts'. The specificity and effectiveness of prompts strongly determine the quality of AI-generated output. Prompt engineering thus has caught huge traction, especially surrounding the use of AI in education settings. Prompt engineering is an  empirical and emerging domain which deals with the ways of systematically structuring clear, concise, and effective prompts that enhance the quality of AI-generated content.  Specifically, prompt engineering refers to the ``{\em systematic design and optimization of input prompts to guide the responses of large language models, ensuring high levels of accuracy, relevance, coherence, and usability in the generated output}''~\cite{chen2025unleashing}. With the growing capability and accessibility of emerging AI models, prompt engineering has been increasingly recognized as an essential competency for the future workforce, particularly in STEM.

Broadly, prompt engineering techniques can be categorized into two types: foundational and advanced. Foundational techniques emphasize clarity, structure, and organization of prompts to optimize outputs. Advanced techniques, on the other hand, involve more sophisticated and dynamic interactions with the responses (such as directing the model `to think step by step' or pretraining it with  disciplinary information) to yield precise, accurate, and high-quality outputs.

In this study, we primarily focus on the foundational techniques and suggest readers~\cite{chen2025unleashing} for the details on advanced techniques. Below, we briefly outline the foundational techniques to better contextualize discussions in the later sections of this manuscript.

\begin{enumerate}
    \item {\bf Providing instructions}. This technique involves designing directives that guide the AI model toward the intended and relevant output. For example, consider the objective of understanding the significance of the resistance $R$ of a substance with respect to temperature ($T$). A prompt such as `{\em Relation between $R$ and $T$}' may yield broad and diverse results with no clear instructions in the statement. Additionally, since $R$ and $T$ could correspond to notations for numerous physical quantities (e.g., Molar gas constant or Time period respectively), the prompt may likely invoke diverse responses.

    \item {\bf Being clear and precise}. Crafting clear and precise instructions helps the model generate accurate and relevant output. In contrast, general instructions often lead to overly broad or ambiguous responses given the model's vast training dataset. For example, if the objective is to gain a basic understanding of electrical resistance, the prompt ``{\em Explain the meaning of $R$ in science}'' may yield broad or contextually irrelevant results.

    \item {\bf Role-based prompting}. The feature of assigning a role to the AI model to yield task-specific output. Within this prompting technique, two subtypes can be distinguished: static role prompting and dynamic role-play prompting. Static role prompting involves assigning a fixed role to the model to enhance contextual accuracy and task-specific performance. Expert Prompting is one such approach, where the AI is explicitly defined as a domain expert. For example, in the context of grading numerical physics problems, static role prompting (aligned with ExpertPrompting) might take the form: ``{\em You are an expert physics instructor tasked with grading student responses...}'' 
    
    Dynamic role-play prompting, in contrast, varies the assigned role to the AI model based on user interaction. Following up with the  grading example above, role-play prompting could extend the static role by supplementing the instruction with: ``{\em Now, expand on the mathematical principle and convey he geometric implications of the results}''. Such dynamic variation is likely to further enhance the accuracy and depth of the output.

    \item {\bf Use of delimiters}. When the task is complex and the prompt has multiple components, delimiters such as triple quotes or custom symbols (*) are used to separate the components. Clear demarcation between the prompt components helps AI generate more coherent and structured output.

    \item {\bf Trying several times}. Also referred to as resampling, this technique involves running the model multiple times and selecting the best candidate output, thereby increasing the likelihood of obtaining an optimal response. It is particularly beneficial given the nondeterministic nature of AI models in generating outputs.

    \item {\bf Zero-shot, One-shot, or Few-shot prompting}. The feature of providing none (Zero-shot), single (One-shot) or multiple examples (Few-shot prompting)  to draw from and generate responses. In the context of zero-shot prompting, the model relies on its large training data for generating responses.

    \item {\bf Variations of model settings}. Varying the parameter settings of the AI model such as `Temperature' can also influence the outputs. The Temperature parameter regulates the degree of randomness of the generated output. A lower temperature parameter can lead to higher deterministic outputs and vice-versa.
\end{enumerate}

Among the seven techniques described above, we adopt five  by omitting the fifth and the seventh from the list. Furthermore, we adopt zero-shot prompting (sixth in the list) by not embedding any examples in our prompt templates. Table~\ref{tab:prompts}, particularly the prompts corresponding to Feedback B and C reflects the five techniques operationalized in our study's context.

\renewcommand{\arraystretch}{1.5}
 \begin{table}[tb]
\begin{ruledtabular}
\caption{The three types of feedback (first column) and corresponding prompt statements (second column) embedding the foundational prompt engineering techniques and principles of effective feedback discussed in Section~\ref{sec:background}.}\label{tab:prompts}

\begin{tabular}{p{0.2\linewidth} p{0.75\linewidth}}

Feedback & Prompt \\
\hline 

Feedback A & [Student generated prompt]. \\ 

Feedback B & ``````You are an expert physicist and your objective is to give feedback on my answer which is presented as an argument with a claim, evidence and reasoning about a physics problem.'''''' \\

& ``````The problem involves identifying the physical significance of a graph’s slope with the gravitational potential energy of a satellite (U) plotted on y-axis and the satellite’s distance from the planet’s center (r) on x-axis.'''''' \\

& ``````The following is my answer (argument): [INSERT YOUR ARGUMENT HERE].'''''' \\

& ``````Provide relevant and useful feedback for my argument.'''''' \\

Feedback C & [Prompt for Feedback B (except for the last sentence)] +  \\

& ``````Provide feedback by explicitly highlighting the correct answer, my provided answer (including strengths and limitations), and gaps (if any) between them within 200 words. Also suggest three potential ways through a bulleted list which can help me improve my performance on similar questions in future.'''''' \\
\end{tabular}
\end{ruledtabular}
\end{table}
 
\subsection{Principles of effective feedback}\label{subsec:effective-feedback}

Feedback is any information provided by an agent (people or material artifacts) focused on aspects of one’s performance~\cite{hattie2007power}. As an integral component of both student learning and teaching practices, feedback serves to bridge the gap between current performance and desired goals. The effectiveness however, depends on the design, content, and structure of the feedback. Effective feedback (student-facing and in the context of formative assessment)  entails the following three components~\cite{hattie2007power,burke2010giving,nicol2006formative}:

\begin{enumerate}
    \item {\bf Desired performance (`{\em Where is the learner going?}')}: This feature involves explicitly and clearly specifying the goals associated with a given task or performance. Clear communication of goals helps students set appropriate self-directed objectives, thereby providing a strong impetus for learning. This aspect of feedback is often referred to as `Feed Up'. For example, in the context of physics problem-solving, this might correspond to clarifying the expected product (e.g., the correct answer) or the process (e.g., use of relevant mathematical formalisms).

    \item {\bf  Current performance (`{\em How is the learner going?}')}:  Clarifying the current state of performance, often in relation to the desired performance. Also referred to as `Feed Back', it includes highlighting both strengths or well-executed parts of the task and gaps in student understanding through probing questions. Such a `praise and question' approach can support students’ self-efficacy, enhance their motivation to engage with feedback, and enable them to develop as self-regulated learners.

    \item {\bf Concrete approaches to fill the gaps (`{\bf Where to next?}')}: Prescriptions about ways to address gaps between current and desired performances. It may include cognitive elements such as task descriptions, specific areas for improvement, actionable suggestions, and, where possible, plans for implementing changes. This component is often referred to as `Feed Forward'.    
\end{enumerate}

\begin{figure}[tb]
    \centering
    \begin{tcolorbox}
     \justify{You are an intern in an astrophysics laboratory, monitoring the motion of a satellite of mass 500$kg$ around a planet (with mass $6.42 \times 10^{23}$ $kg$) in a nearby galaxy. You receive data the following from the observatory highlighting the satellite’s gravitational potential energy with respect to its distance from the planet’s center.\\}

\noindent
\begin{tabular}{p{0.28\linewidth} p{0.28\linewidth} p{0.4\linewidth}}
\hline
  {Altitude `$h$'} & Distance `$r$' & Potential energy `U' \\
  (km) & (m) $\times 10^6$ & (J) \\
\hline 
  100 & $3.49$ & $ -6.11$ \\
  200 & $3.59$ & $ -5.94$ \\
  300 & $3.69$ & $ -5.78$ \\
  400 & $3.79$ & $ -5.62$ \\
  500 & $3.89$ & $ -5.48$ \\
\end{tabular}

\justify{(1). Based on the provided data, produce a detailed written solution (including the concepts, principles, assumptions and approximations, and your strategy) to determine the values of the planet’s:

(a) Radius

(b) Magnitude of acceleration due to gravity on its surface.}

\justify{Submit your detailed written solution in a pdf or image file format.}

\justify{(2). Consider a plot of $U$ ($y-$axis) vs $r$ ($x-$axis). What would be the physical significance of the graph’s slope? Detail your argument by explicitly highlighting your Claim, Evidence, and Reasoning using relevant equations, diagrams, and physics principles.}

\end{tcolorbox}
\caption{Statement of the entire task. Problem 2 was the specific task for which students sought feedback from AI.}\label{fig:problem-statement}
\end{figure}

In addition to these components, the timing of feedback plays a crucial role in its effectiveness. Immediate task-related feedback can be beneficial for simpler tasks, whereas delayed, process-oriented feedback may be more effective for complex or challenging tasks.

In the present study, one of the prompt templates provided to students for eliciting feedback from AI (`Feedback C' in Table~\ref{tab:prompts}) incorporated the three components described above. Furthermore, the AI-generated real-time feedback appropriately addressed the timing factor.

\section{Methods}\label{sec:methods}

\subsection{Data and Context}\label{subsec:data-context}

Our data comes from introductory students’ responses to an extra-credit assignment at a Midwestern land-grant R1 university in the United States. The assignment was part of a large-enrollment calculus-based physics course for future engineers and was administered through Qualtrics~\footnote{https://www.qualtrics.com/} during Spring 2025. The course follows the Matter and Interactions textbook by Chabay and Sherwood~\cite{chabay2015matter}, with a focus on the principles of energy, momentum, and angular momentum. It also includes assignments in which students constructed arguments using the `Claim, Evidence, and Reasoning (CER)' structure~\cite{toulmin2003uses,mcneill2008inquiry}.

The assignment focused on the variation of a satellite's potential energy with respect to its distance from a planet’s center. Hypothetical data corresponding to the satellite’s altitude from the planet’s surface, its distance from the planet’s center, and associated  potential energy were provided in a table (Fig~\ref{fig:problem-statement}). The task  consisted of three parts. The first part focused on determining the planet’s radius and magnitude of the acceleration due to gravity on its surface. Students were asked to submit their written solutions in PDF or image file formats. The task's second part focused on interpreting the physical significance of the slope of a graph with gravitational potential energy plotted against the satellite’s distance from its center. Students were asked to communicate this significance through arguments structured in the form of Claim, Evidence, and Reasoning (CER). 

The final part of the assignment required students to seek feedback from AI platforms of their choice on their arguments about the graph’s slope (second task). The task was intentionally designed to encourage students to include relevant details in their prompts rather than copying-and-pasting the problem statement. This allowed to gather data on how students constructed prompts which would further be analyzed through prompt engineering techniques. Students interacted with AI through three prompt types: (i) self-crafted prompts, (ii) a provided prompt incorporating foundational prompting techniques (Section~\ref{subsec:prompt-engineering}), and (iii) a provided prompt combining foundational prompting techniques with principles of effective feedback (Section~\ref{subsec:effective-feedback}). The corresponding feedback from AI to these three prompts are henceforth referred to as `Feedback A', `Feedback B', and `Feedback C'. For the prompts that were provided,  students were required to insert their arguments into highlighted areas. No specific AI platform was recommended, allowing students to turn to the platform of their choice and enabling more authentic observations of their interactions. The survey did not allow  students to go back and modify their prompts or their responses at any point during the task.

Information regarding both their self-crafted and provided prompts along with corresponding responses (feedback statements) from AI were collected.  In the final stage of the task, participants were asked to rank the three feedback types by perceived usefulness and provide detailed justifications. They were also asked to list up to three things (if any) they had learned from the activity and to reflect on its usefulness for their learning. Our analysis focuses primarily on this third part of the task i.e., students’ engagement with AI feedback, particularly on student-generated prompts, their rankings of feedback, and their justifications.

\subsection{Analysis}\label{subsec:analysis}

Of the 2044 students enrolled in the course during the Spring 2025 semester, 1661 responded to the extra-credit activity (81\% response rate). The data were downloaded into a spreadsheet, and incomplete responses were removed, resulting in 1235 qualified responses. To address the first research question, students’ self-crafted prompts were analyzed through the lens of foundational techniques, with a particular focus on the first four techniques (`Providing instructions', `Being clear and precise', `Role-based prompting' and `Use of delimiters'). The remaining three techniques: `trying several times’, `one-shot/few-shot prompting’, and `variation of model settings’ were not included either in the provided prompt design or in the analysis of student-generated prompts.

The first four techniques were coded dichotomously (`y' or `n'), with `y' indicating presence and `n' indicating absence of evidence of foundational techniques. Prompts that included image or PDF attachments (17.6\%) were reviewed and coded separately. Evidence of the first technique, `providing instructions’ was coded as present if students’ prompts included expectations from the AI model, such as seeking feedback or evaluating their arguments. Any instruction or request in students’ prompts about their arguments was considered evidence for this technique. The second technique, `being clear and precise’ was coded into two subcategories: (i) explicit description of the task’s objective as identifying the slope of a graph of gravitational potential energy versus distance from the planet’s center, and (ii) explicit mention of the student’s argument or their final claim.  Both subcategories were coded separately on a binary scale (`y' or `n') and only if students evidenced both the components, the prompt was coded as evidencing the second technique. Evidence for the third technique, `role-based prompting’ was coded as present if the prompt assigned a role to the AI model or else was marked as absent. In a similar fashion, the evidence for the final technique, `use of delimiters' was also coded. A total of 84 student prompts (around 7\%) were excluded since they were either unanswered or sought feedback for earlier parts of the task.

For the second research question, the first author employed emergent coding approach~\cite{creswell2007qualitative} to analyze students’ justifications for their preferred feedback types. Justifications that were left unanswered or were difficult to interpret were removed, leaving 1160 valid responses from the initial 1235 (75 disqualified). These responses were analyzed in two stages. In the first stage, emergent codes were identified by noting the features that students referred to while justifying their choice. A `code sheet’ was then created documenting each feature along with its corresponding meaning (Table~\ref{tab:feedback-features}).  In the second stage, each student justification was coded in binary (`yes' or `no') based on the features identified in the first stage to capture their prevalence across the dataset. Two independent researchers reviewed the code-book and coded a random subset of 25 responses. Their coding was compared and discussed until full agreement was reached. No major additions or deletions were made to the original codes identified in the first stage after the discussion. For the third research question, basic spreadsheet functions such as `COUNTIF’ were used to identify trends in students’ rankings of the three feedback types and the associated feedback features identified earlier.

\renewcommand{\arraystretch}{1.5}
 \begin{table*}[tb]
\begin{ruledtabular}
\caption{Exemplar student-crafted prompts reflecting the foundational prompt engineering techniques. Section of the prompt highlighting the technique has been underlined.}\label{tab:exemplar-prompts}

\begin{tabular}{p{0.2\linewidth} p{0.8\linewidth}}

Prompting technique & Example Prompt \\
\hline 

Providing Instructions & {\em The slope of a graph (U vs r, gravitational potential energy vs distance) is gravitational force because newton's universal gravitation law tells us that $F = GMm/r^2$ and the derivative $U$ with respect to $r$ gives us the same equation. \uline{Can you give me feedback on this claim?}}  \\ 

Being Clear and Precise & {\em Hey! I’m working on a physics assignment and had to \uline{explain what the slope of a gravitational potential energy vs. distance (U vs. r) graph means}. I wrote my answer using a Claim, Evidence, and Reasoning format. Can you let me know if it makes sense and if I explained the idea clearly?

\uline{Claim:} [I think the slope of the graph shows ....]

\uline{Evidence:} [From the data, I saw that...]

\uline{Reasoning:} [If the slope is steep, it means ....] }\\ 

Role-based prompting & {\em \uline{You are a specialized physicist} on the gravitational energy field. I had a task about finding the physical significance of the slope in a plot that compared the gravitational potential energy (U) in the y-axis, with the radius (r) in the x-axis. I need you to give me brief feedback (keep it to a maximum of 200-300 words) of my response to this question. My response has been attached.}  \\ 

Use of Delimiters & {\em Hey ChatGPT! Based on the following definition of claim, evidence, and reasoning, please give feedback about the argument written in claim evidence and reasoning format!

\uline{The definitions of claim, evidence, and reasoning will be pasted between asterisks (*) and the argument presented by a student will be pasted between quotations (")}.

Please give feedback based both on the validity of the argument and the adherence to the formatting of claim, evidence, and reasoning.

Thanks!

\uline{*}A Claim is every decision [...]\uline{*}

\uline{``}The slope of GPE vs r represents [...]\uline{''}} \\

\end{tabular}
\end{ruledtabular}
\end{table*}

\begin{figure}
    \centering    
    \includegraphics[width=0.48\textwidth]{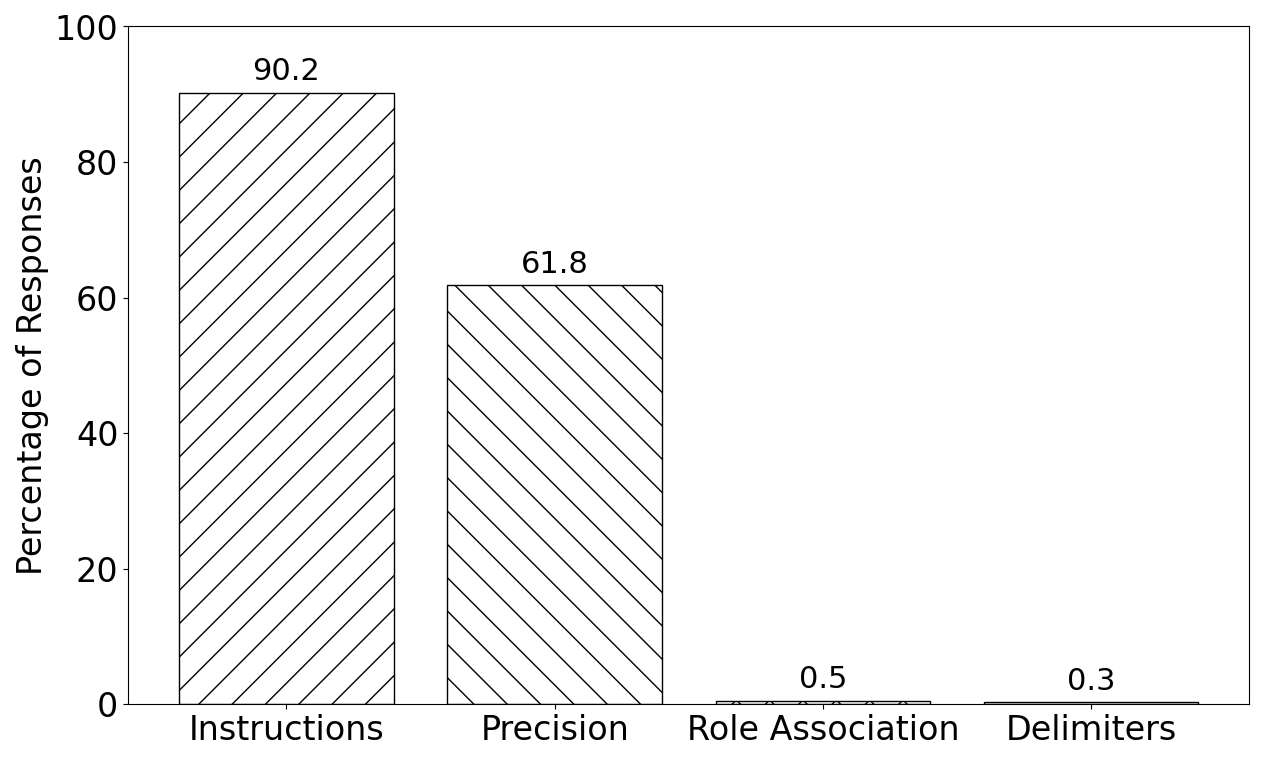}
    \caption{Graph highlighting the percentage of student-crafted prompts reflecting the foundational prompt engineering techniques. Each of the four columns correspondingly represent the following four prompting techniques: `Providing instructions', `Being clear and precise', `Role-based prompting' and `Use of delimiters'. 6.8\% of student responses were rejected (not highlighted in the graph).}\label{fig:prompt-sophistication}
\end{figure}

\section{Results}\label{sec:results}
In the rest of this section, we detail the results of the three research questions outlined in Section~\ref{sec:intro}. The first question examines the extent of sophistication evidenced in student generated prompts through the lens of foundational prompt engineering techniques (Section~\ref{subsec:prompt-engineering}). The second question focuses on the valued features students attributed to the perceived usefulness of AI-generated feedback. The last question explores the trends in students’ ranking of the feedback generated from: (i) self-crafted prompts, (ii) provided prompts embedding foundational prompt engineering techniques, and (iii) provided prompts entailing the combination of foundational techniques along with the features of effective feedback (Section~\ref{subsec:effective-feedback}). The following subsections explore each research question.

\begin{figure}
    \centering        
    \includegraphics[width=0.48\textwidth]{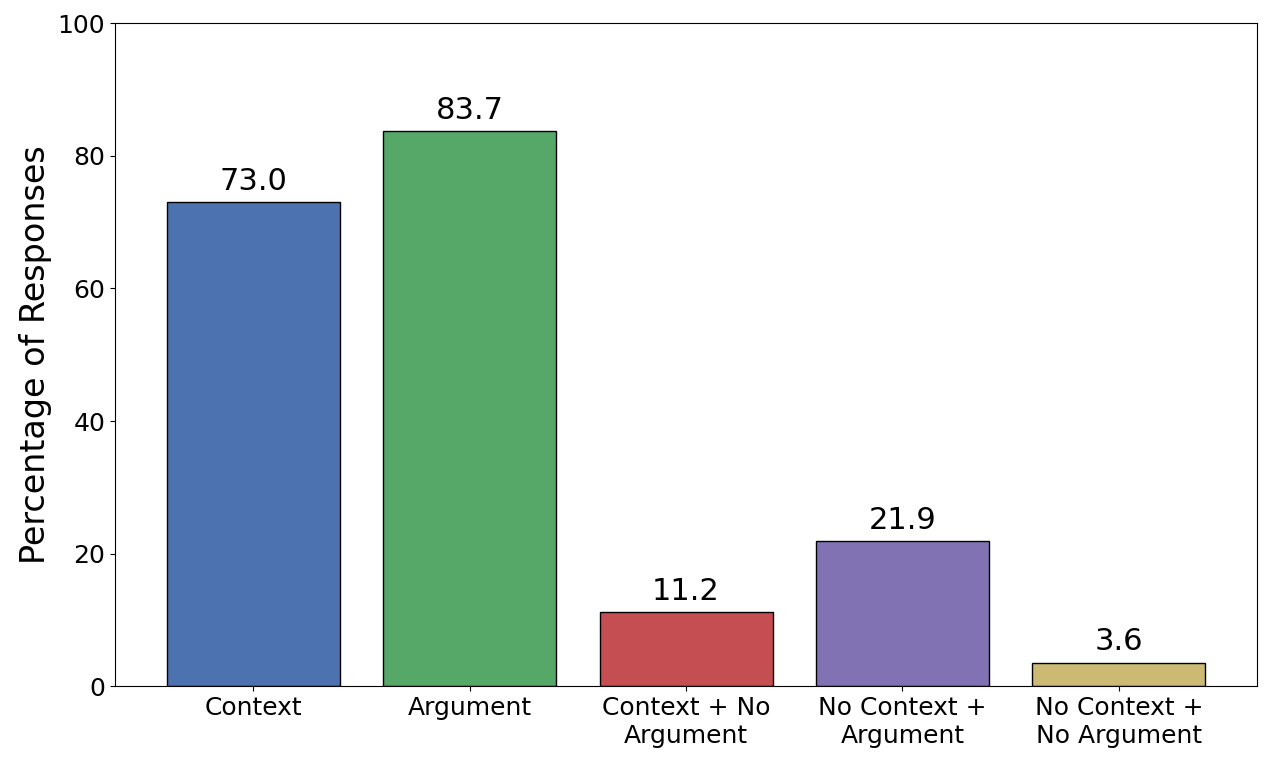}
    \caption{Graph highlighting the details of student-crafted prompts reflecting the foundational technique of ``Being Clear and Precise''. This technique was coded when students described the problem's context and accompanied their arguments.}\label{fig:prompt-clarity-and-precision.png}
\end{figure}

\subsection{Prompt sophistication} 

{\em {\bf RQ1: }To what extent do students' self-crafted prompts to AI evidence the features of foundational prompt engineering techniques?}

As discussed in Section~\ref{subsec:data-context}, we evaluated students’ self-crafted prompts in terms of prompt-engineering techniques, focusing on `Providing clear instructions', `Being clear and precise', `Role-based prompting', and `Using delimiters'. The distribution of these features is presented in Figures~\ref{fig:prompt-sophistication}~and~\ref{fig:prompt-clarity-and-precision.png}, while Table~\ref{tab:exemplar-prompts}  provides exemplar prompts evidencing each technique.

The results reveal three overarching trends: (i) none of the students’ prompts  contained all four foundational techniques together, (ii) contemporary AI models often demonstrated sophistication in interpreting contexts when prompts were ambiguous, and (iii) despite this sophistication, AI responses at times did not provide appropriate feedback.

\subsubsection{Providing clear instructions}

Overall, 90.2\% responses evidenced the technique of providing clear instructions in their prompts by explicitly seeking feedback for their arguments. These instructions were framed as requests (``{\em Please review and provide feedback for the argument}''), commands (``{\em Give me feedback to this argument'')}, or questions (``{\em Can you let me know if it makes sense and if I explained the idea clearly?}''). The remaining 9.8\% prompts  provided only arguments (with no explicit instruction to provide feedback) and yet the model consistently responded with feedback. For instance, for one of the students’ prompt with no explicit instruction for feedback:

\begin{quote}
    {\em The slope of the gravitational potential energy U vs r graph is equal to the negative of the gravitational force acting on the satellite. From the equation U = -GMm/r the derivative is dU/dr = GMm/$r^2$. The slope of the U vs r graph tells us how fast potential energy changes with distance. The force is the negative rate of change of potential energy. The slope of this graph represents gravitational force.}
\end{quote}

The AI responded with feedback: 

\begin{quote}
    {\em Your argument demonstrates a strong conceptual understanding of the relationship between gravitational potential energy and gravitational force. You correctly identified that the slope of the U vs r graph represents the derivative and you clearly linked it to the gravitational force.}
\end{quote}

\subsubsection{Being clear and precise}

To capture the second technique, `Being clear and precise', we analyzed students' prompts across two components: (i) explicit description of the task’s context by specifying the slope of gravitational potential energy versus distance graph and (ii) explicit mention of arguments. If students evidenced both  components, we coded the prompt as evidencing this technique. While 73\% of students clearly described the graph's context and 83.7\% included their arguments, only 61.8\% demonstrated both together. In contrast, 11.2\% provided context without arguments, 21.9\% included arguments with unclear context, and 3.3\% provided neither. These trends are represented in Fig.~\ref{fig:prompt-clarity-and-precision.png}.

Given these variations, we further examined how AI responses varied when prompts lacked clear or complete context descriptions. Of the 293 such cases, AI correctly interpreted the context and generated feedback in 285 instances. In only 8 cases did AI request additional clarifications or recommend significant rewording of the prompt. For example, for one student’s prompt with incomplete context regarding the type of energy (potential vs. gravitational potential):

\begin{quote}
    “{\em Provide me with feedback about my claim, evidence, and reasoning regarding the potential energy and distance relationship. Claim: As the distance increases, the potential energy increases. Evidence: There is a positive slope between the distance and potential energy. Reasoning: Therefore there is a positive and linear relationship between potential energy and distance. However, the magnitude of the potential energy decreases as distance increases.}”
\end{quote}

AI responded by suggesting clarity in the argument: 

\begin{quote}
“{\em Your claim, evidence, and reasoning contain important observations, but there are a couple of points that could be clarified for better consistency and accuracy. Here is some feedback:

Claim: The statement ``As the distance increases, the potential energy increases'' is straightforward, but it assumes a specific scenario. Clarify whether this refers to gravitational potential energy, elastic potential energy, or another type of potential energy, as the relationship can vary depending on the context.}''
\end{quote}

Furthermore, though AI demonstrated impressive `contextual intuitiveness' (i.e., the feature of appropriately capturing the right context despite contextually ambiguous prompts) in 97\% of cases, these responses were not always fully appropriate or correct. For instance, in one case a student described the slope of the graph as representing the variation of a generic potential energy with distance, without specifying gravitational potential energy or reasoning about the slope as gravitational force. The AI’s response merely corrected the relationship between the quantities but did not prompt further clarification or address the incorrect physical interpretation of the slope:

Prompt:

\begin{quote}
    ``{\em I was given some data which I graphed and was asked to find the physical significance of the slope of potential energy vs distance graph. I graphed it and the data had a positive slope. I argued that the positive slope indicates that the gravitational potential energy decreases as the distance between the planet and satellite increases. What do you think of my argument?}''
\end{quote}

AI response:

\begin{quote}
``{\em Correct understanding: gravitational potential energy changes with distance.

Correction needed: potential energy increases (not decreases) with distance, even though the values are negative—you're moving toward zero.

Let me know if you want help connecting this to orbital motion or force interpretation!}''
\end{quote}

\subsubsection{Role-based prompting and Use of delimiters}

The third and fourth techniques (`Role-based prompting' and the `Use of delimiters') were the least evidenced in students' prompts. Out of 1151 responses, only six demonstrated role-based prompting, and only three included explicit use of delimiters such as quotes and asterisks. Exemplar prompts illustrating these two techniques are presented in Table~\ref{tab:exemplar-prompts}.

\renewcommand{\arraystretch}{1.2}
 \begin{table*}[tb]
\begin{ruledtabular}
\caption{Themes and corresponding features associated with the features of feedback evidenced in students' descriptions of their preferred feedback type. Themes are highlighted in the bold text. Numbers in the parenthesis represent the percentage of student responses which evidenced at least one of the associated features.}\label{tab:feedback-features}

\begin{tabular}{p{0.2\linewidth} p{0.78\linewidth}}

Theme/Code & Description \\
\hline 

\textbf{Evaluation (69.4\%)}    & \textbf{Strengths and shortcomings in arguments OR Suggestions for improvement}  \\

E1: Affirmation    &  Appreciation or acknowledgment of the correct sections of the argument \\

E2: Critique    &   Pointing out errors or miscalculations in the argument \\

E3: Guidance    &   Suggestions for improvement OR Pointing to specific areas of deliberate practice \\

\textbf{Content (55.2\%)} & \textbf{Exemplar correct answers OR Physics/mathematical principles OR Formulae OR Generic argument OR Argument in the form of Claim, Evidence, Reasoning}g \\ 

C1: Correctness    & Exemplar “correct” solutions OR ``The right approach''  \\

C2: Conceptual     &  Physical principles or mathematical relations \\

C3: Schema         & Focus on argument or on Claim, Evidence, and Reasoning structure \\

\textbf{Presentation (47.8\%)}    & \textbf{Feedback’s Clarity OR Structured formatting OR Brevity OR Ease of understanding}  \\

P1: Clarity   &  Clear communication of feedback information OR Description of the task \\

P2: Structure &   Well-formatted structure of feedback OR Feedback with distinct sections    \\

P3: Conciseness &   Less on “wordiness” but more on useful information    \\

P4: Intelligibility  &   Easy-to-understand OR Simple phrasing    \\

\textbf{Depth (44.2\%) }   &  \textbf{Detailed description OR detailed solution OR Task-related specificity of feedback}  \\

D1: Detail    & Detailed account of student solution OR Detailed feedback  \\

D2: Specificity    & Task-specific feedback OR Feedback with strong focus on student solution  \\

\end{tabular}
\end{ruledtabular}
\end{table*}

\subsection{Valued features in AI-generated feedback}

{\em {\bf RQ2:} What features of AI-generated feedback do students perceive as useful when the feedback is based on (a) self-crafted prompts, (b) prompts incorporating foundational prompt-engineering techniques, and (c) prompts combining foundational prompt-engineering techniques with principles of effective feedback?}

Based on the emergent coding of students’ justifications of their preferred feedback types (generated from the three different prompts), we identify the following four themes associated with students' valued features in AI-generated feedback: (i) Evaluation, (ii) Content, (iii) Presentation, and (iv) Depth. Table~\ref{tab:feedback-features} presents the description of each theme along with the associated features. Table~\ref{tab:feedback-feature-example} provides exemplar student quotes that reflect each feature. Below, we describe each theme and associated features.

\subsubsection{Evaluation (E)}

The first predominant theme corresponds to features of feedback that critically evaluate the strengths and shortcomings of student solutions along with a clear description of ways for improvement to facilitate better performance on similar tasks in future. Three features are associated with this theme: (E1) Affirmation (acknowledgment of the solution’s correctness or strengths), (E2) Critique (pointing out errors or shortcomings of the solutions) and (E3) Guidance (actionable suggestions for improvements). A total of 65.2\% of student descriptions evidenced at least one of these features in their justifications. The following is one of the quotes which evidences all three features:

\begin{quote}
    
{\em I found Feedback B the most useful because it gave a clear evaluation of my argument and highlighted areas of improvement without being vague. It confirmed what I did well (identifying the slope as the negative of the force) and then pointed out small gaps, like not fully explaining the curvature of the graph and not connecting the shape to the inverse-square nature of gravity. This helped me reflect more deeply on physics, not just the formula. I also appreciated the suggestion to use real-world examples, which would make my explanation stronger and more relatable.}
\end{quote}

The student’s reference to feedback acknowledging what was done well, i.e., identifying the graph’s slope as the negative of the gravitational force between the planet and satellite, illustrates the `Affirmation' feature. In addition, the feedback’s identification of minor gaps in the solution, along with recommendations to include real-world examples in the arguments, reflects the `Critique' and `Guidance' features, respectively.

\subsubsection{Content (C)}

This theme reflects feedback features that predominantly focus on content knowledge. These include the elucidation of the `correct' approach to solving the physics problem or the specification of the `correct answer'. The theme also encompasses the application of physics and mathematics principles, along with relevant equations. In addition, it involves explicit discussion of content knowledge embedded within the argumentation structure, i.e., claim, evidence, and reasoning. We identified three features associated with this theme: (C1) Correctness (focus on the correct answer), (C2) Conceptual (discussion of physics or mathematics principles), and (C3) Schema (integration of content knowledge within the argument structure). Overall, 51.8\% of responses evidenced at least one of these features. The following quote illustrates all three features associated with the Content theme:

\begin{quote}
    {\em Feedback C was the most helpful because it provided mathematical explanations with clear equations illustrating any mistakes in my argument. Additionally, it explicitly highlighted the correct answer and allowed me to compare it to my answer, easily allowing me to understand any flaws. Lastly, it provided a list of tips to improve my argument, which was far more useful than the other two arguments, which only provided at most 1-2 areas for improvement.}
\end{quote}

The reference to `{\em explicitly highlighting the correct answer}’ evidences the Correctness feature. In addition, the inclusion of mathematical explanations with clear equations reflects the Conceptual feature, while the use of an enumerated list with recommendations to strengthen the argument illustrates the Schema feature.  

\renewcommand{\arraystretch}{1.3}
 \begin{table*}[tb]
\begin{ruledtabular}
\caption{Exemplar student responses highlighting each feedback feature. Underlined text in the student responses reflect the corresponding feature.}\label{tab:feedback-feature-example} 

\begin{tabular}{p{0.14\linewidth} p{0.84\linewidth}}

Code & Example \\
\hline 

E1: Affirmation    & {\em I found the third feedback to be the most useful. [...]  It \uline{allowed me to see what I did well and what I should continue to do.}} \\

E2: Critique    & {\em Feedback C, as it actually \uline{pointed out that I was wrong}, others felt like they were half-hearted explanations and were just glazing my answer but feedback C actually \uline{pointed out I was wrong} which made me feel more eager to learn my mistake even though I felt that all 3 of the answers I put in, I felt unclear about all but only feedback C made me \underline{realise the error} and gave me a better answer from which I could learn off.}   \\

E3: Guidance    & {\em Feedback C was the most useful because it \uline{gave me ways to improve in the future} instead of just correcting what I did wrong like A and B.} [...]   \\

C1: Correctness    & {\em A is the most useful because \uline{it provides the right answer}. The other 2 say that we got a negative slope, which is wrong.}  \\

C2: Conceptual     &  {\em I liked Feedback B the most and found it most useful because it \uline{helped me understand the overall shape and behavior of the gravitational potential energy vs. distance graph}. By mentioning that the curve is hyperbolic and that the slope decreases due to the inverse square law. It gave me a clear visual and conceptual understanding of how gravity weakens with distance. This made it easier for me to interpret the graph and connect it to real-world gravitational behavior, rather than just thinking in terms of equations. It deepened my conceptual grasp, not just my ability to calculate.} \\

C3: Schema         &  {\em C was the most helpful because it \uline{specifically highlighted which parts of my arguments} were strong and which weren't.} \\

P1: Clarity   & {\em I found B the most useful. It provided helpful feedback and had a lot of explanation. \uline{It was clear}.} [...]  \\

P2: Structure & {\em I liked B the best because to me it seemed to make the most sense it \uline{organized its response} in a way that made sense to me.} [...] \\

P3: Conciseness & {\em I found C the most useful because it felt that \uline{it was the most concise}.} [...]   \\

P4: Intelligibility  &  {\em I liked A the most as it was much more \uline{easier to understand} compared to the others.}   \\

D1: Detail    & {\em I found the third and last feedback to be the most effective and useful since it went more in depth about all of the information that I needed.}  \\

D2: Specificity    & {\em I found C the most useful because it \uline{did not really run in circles}, the rest kind of reiterated the same point over and over again, but c \uline{was straight to the point}.} \\

\end{tabular}
\end{ruledtabular}
\end{table*}

\subsubsection{Presentation (P)}

The third theme encompasses features related to the presentation of feedback that are independent of the task’s content. These include the clarity of the provided information, the external structure or formatting (e.g., use of an enumerated list), the feedback’s length or brevity, and its overall ease of comprehension. Four features capture this theme: (P1) Clarity, (P2) Structure, (P3) Conciseness, and (P4) Intelligibility. Approximately 45\% of student justifications referenced at least one of these features. The following quote illustrates all four features associated with this theme:

\begin{quote}
{\em I found Feedback C to be the most useful and my favorite among the three because it offered a clear, structured, and detailed evaluation of my argument. What stood out was how it explicitly separated the correct answer, my response, and any gaps between the two. This side-by-side comparison made it easy to understand not only what I did well but also what specific areas could be improved. Additionally, the three bulleted suggestions at the end were especially helpful because they were clear, actionable, and directly applicable to future problems. The word limit also made the feedback feel focused and concise without losing depth. [..]}
\end{quote}

The response’s repeated references to clarity of evaluation of the argument (first sentence) and actionable suggestions (fourth sentence) evidence the `Clarity’ feature. Along the same lines, the response’s multiple references to structured evaluation of the argument, i.e., `side-by-side’ comparison of what was expected and what was written along with bulleted suggestions reflect the `Structure’ feature of the feedback. The last sentence on the feedback being short yet specific, evidences the `Conciseness’ feature. Lastly, the feedback's ease of comprehension, the strengths and weaknesses of the argument mark the final `Intelligibility’ feature.

\subsubsection{Depth (D)}

The last theme corresponds to the extent of detailed discussion of the student’s solution within or beyond the purview of the task. This theme however does not correspond to the `quantity’ or `wordiness’ of the feedback but rather the quality of detailed discussion of the students’ solution. Two features are associated with this theme: (D1) Detail (detailed discussion of the solution) and (D2) Specificity (feedback’s primary focus on the solution with reference to the task with no generic discussions). 44.2\% of students’ descriptions evidenced at least of these two features. An example highlighting the four features is provided below. 

\begin{quote}    
{\em I think that [Feedback] B gave more direction in the best way to explain the prompt in a detailed manner. It gave specific answers as to what was missing from my response and how to improve it, where as the others were more generalized.}
\end{quote}

In the above quote, the first sentence referring to the feedback feature to explain the task in a detailed fashion evidences the `Detail’ feature. On the other hand, the next sentence referring to feedback’s specificity about the shortcomings and ways to improve the solution mark the second feature, i.e., `Specificity’.  

\begin{figure}
    \centering
    \includegraphics[width=\linewidth]{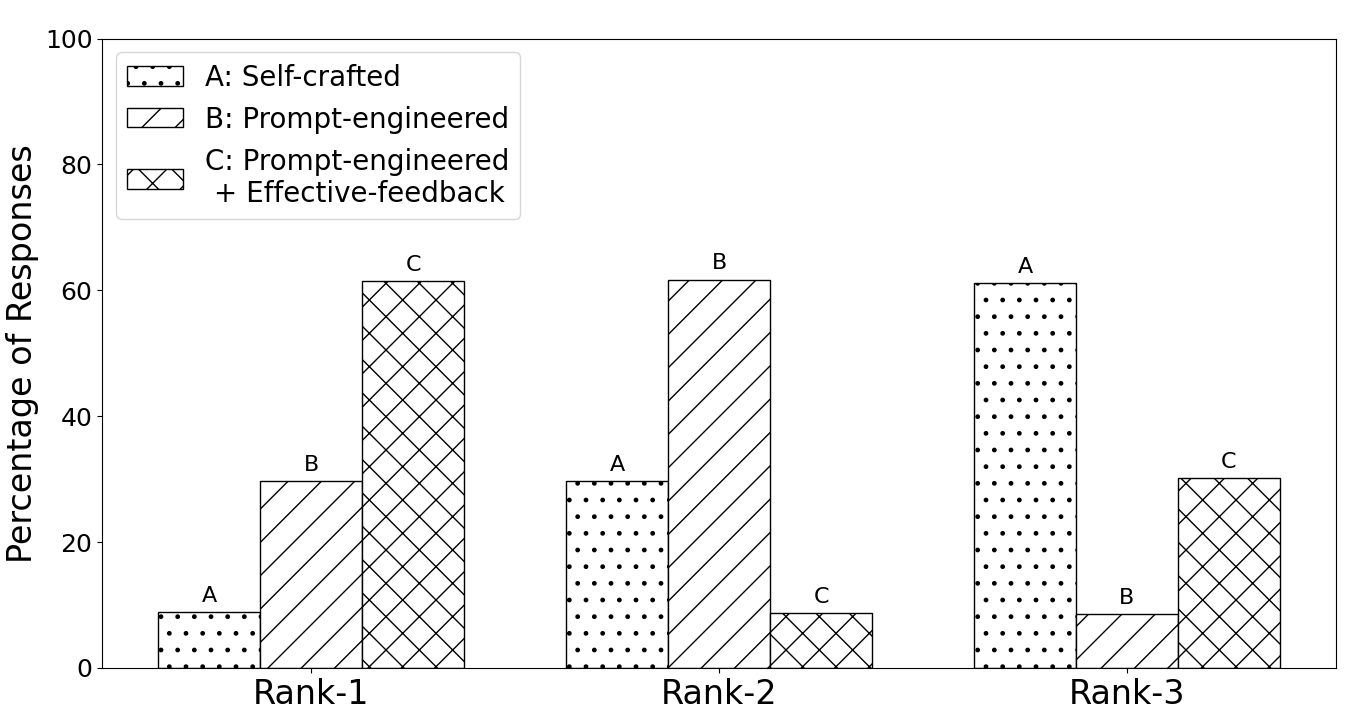}
    \caption{Plot highlighting students' ranking of the three types of feedback generated from prompts that: (i) were self-crafted, (ii) entailed features of prompt engineering, and (iii) entailed features of prompt engineering along with principles of effective feedback.}\label{fig:ranking}
\end{figure}

\subsection{Student preferences of AI-generated feedback across prompt types}

\subsubsection{Feedback Ranking}

 {\em {\bf RQ3:} What are the broad trends in students’ preferences for AI-generated feedback based on (a) their self-crafted prompts, (b) prompts entailing foundational prompt-engineering techniques, and (c) prompts combining the foundational prompt-engineering techniques with principles of effective feedback?}

As noted in Section~\ref{sec:methods}, students used AI models of their choice to generate three distinct feedback statements through three different prompts: (i) self-crafted prompts (resultant feedback referred to as `Feedback A’), (ii) provided prompts that entailed features of foundational techniques of prompt engineering (Feedback B), and (iii) provided prompts that entailed the combination of foundational techniques and the principles of effective feedback (Feedback C). Students then ranked the three feedback statements generated by AI based on their perceived usefulness along with relevant justifications. Here, we discuss the trends in students' ranking of the three feedback types along with the trends associated with students' attribution of the feedback features (discussed in the previous subsection) across each feedback type.

Figure~\ref{fig:ranking} highlights the trends in students' preferences of the three feedback types. Overall, students preferred Feedback C (generated from the provided prompt having the features of both prompt engineering and effective feedback) as their top choice and rated Feedback A (generated from self-crafted prompts) as their least preferred one. Thus, an overwhelming majority of students preferred feedback generated from structured prompts (combined preferences across Feedback types B and C). These trends are detailed below.

Across 1235 responses, around 61.4\%  of students ranked Feedback C as the most useful, followed by B (29.7\%) and A (8.9\%). For their second choice, 61.7\% chose Feedback B, followed by A (29.6\%), and  C (8.7\%). Finally as their least preferred feedback, a major section of students ranked Feedback A (61.2\%), then C (30.2\%), and lastly B (8.6\%). These results highlight three overarching trends about student preferences. Firstly, a substantial section (91.1\%) of students preferred feedback generated using structured prompts (combined first choices across B and C). Within this preference, a majority of students preferred the feedback generated using the prompt with combined features of prompt engineering and effective feedback (C). Secondly, this top choice was either liked or disliked by students. This is evident from 61.7\% indicating C as their top choice and 30.1\% preferring it as their least one. Only 8.2\% however highlighted C as their second choice. Lastly, students' least preferred feedback corresponded to the one generated using their self-crafted prompts.

\begin{figure}
    \centering
    \includegraphics[width=\linewidth]{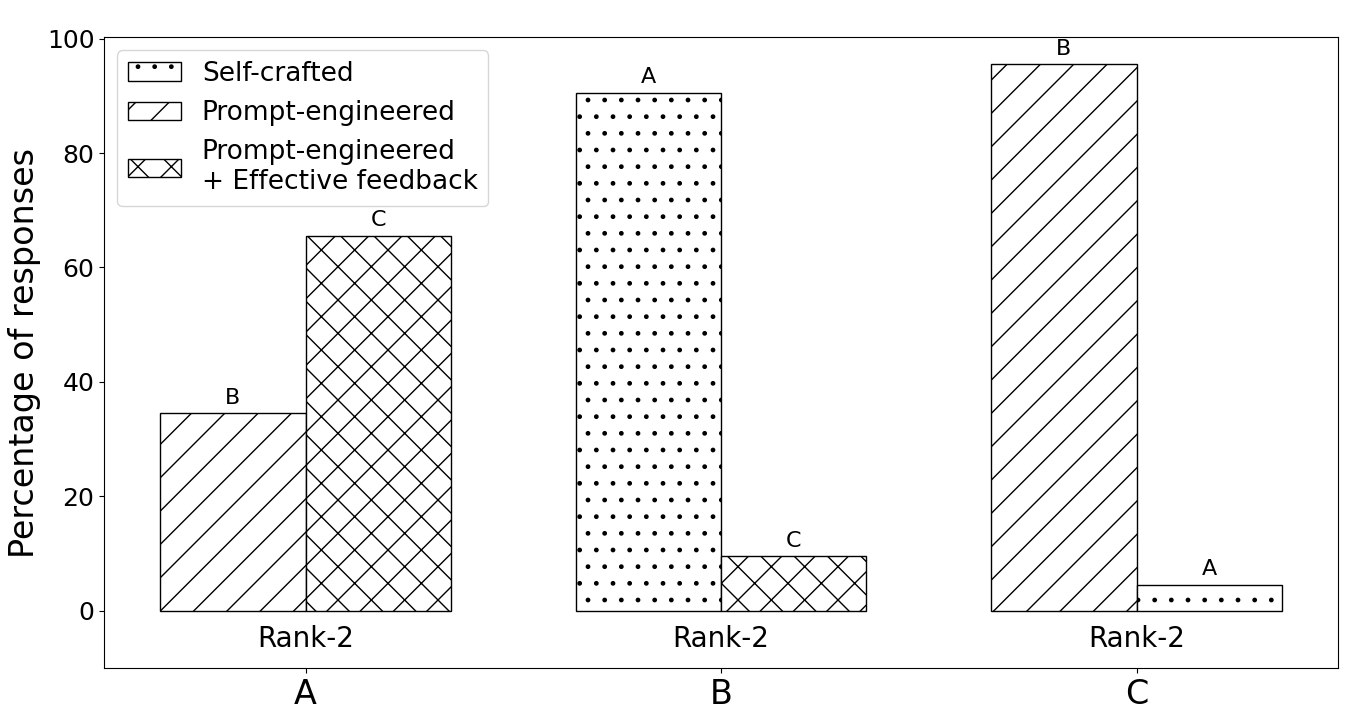}
    \caption{Plot highlighting students' second choices given their first choice of each feedback type. The feedback types at the center of the two columns on the x-axis represent the preferred feedback choice whereas those on top of the columns indicate the second choice.}\label{fig:rank-distribution}
\end{figure} 

Beyond these overarching trends, we also investigated students’ patterns in their second choice (rank) given their first-choice of feedback type. Figure~\ref{fig:rank-distribution} illustrates these trends. Among students who ranked  A as their top choice, two-thirds chose C as their next preferred version with remaining one third choosing B.  An overwhelming 90\% of students who ranked Feedback B as their top choice, selected A, as their second preference. The remaining 10\% however chose C. Finally, 95.5\% of students who preferred Feedback C, selected B as their second choice followed by mere 4.5\% for A.  

\begin{figure*}
    \centering
    \includegraphics[width=\linewidth]{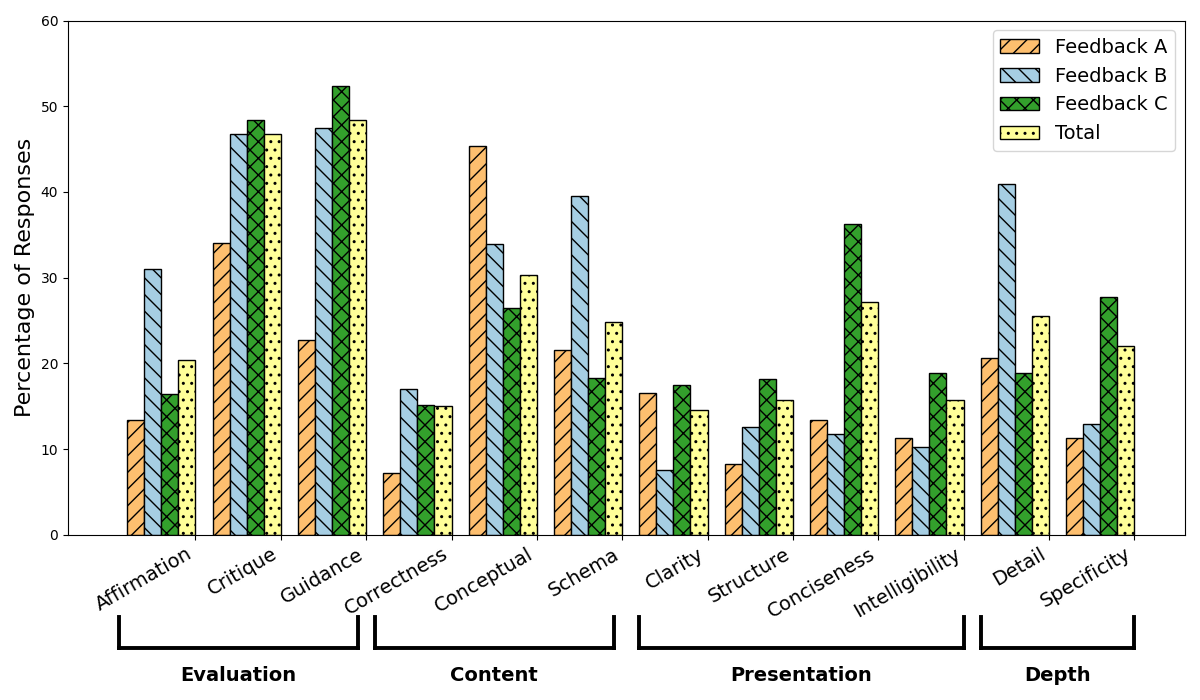}
    \caption{Graph highlighting the percentage of responses of the total data and that across the three feedback types in terms of the feedback features. Associated themes across all features have been highlighted at the bottom end with bold text.}\label{fig:combined-all-features}
\end{figure*}

\subsubsection{Themes and features across the feedback types}

\begin{figure}
    \centering
    \includegraphics[width=0.48\textwidth]{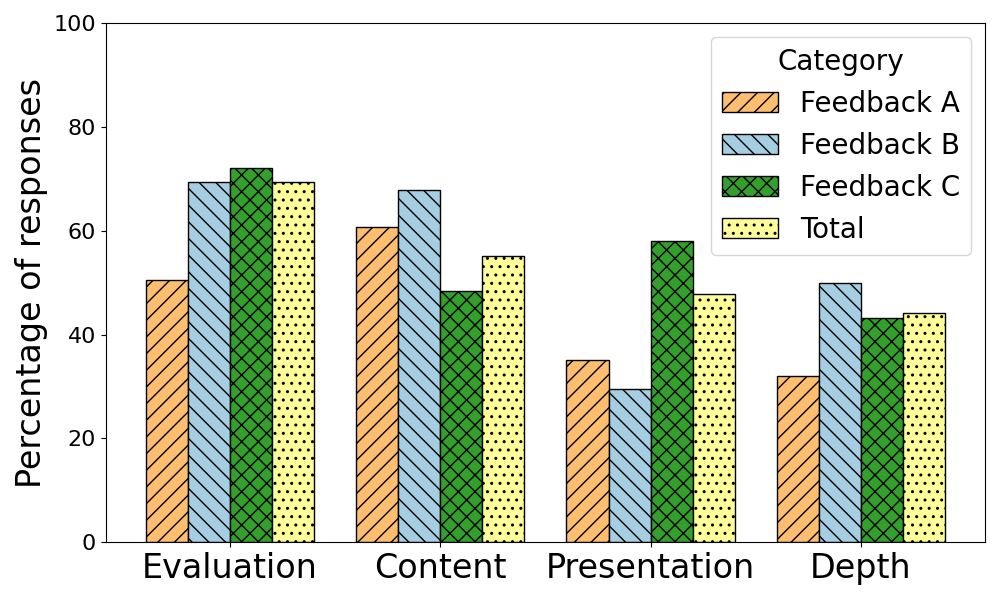}
    \caption{Graph highlighting the percentage of responses of the total data and that across the three feedback types in terms of the feedback themes.}\label{fig:themes-combined}
\end{figure}

In addition to investigating students’ preferred choice of feedback type, we also explored feedback features identified in the previous subsection (RQ2) that students attribute across each feedback type. Fig~\ref{fig:combined-all-features} presents the trends among the 12 feedback features across the three feedback types as well as the entire data. Fig~\ref{fig:themes-combined} presents trends across the four themes. Below, we provide the summary of the results before discussing them in detail.

Across the entire data, students mostly the valued feedback that identified inconsistencies in their solutions and provided actionable recommendations for improvement. This was followed by a preference for content knowledge, particularly in the context of the given task. Concise communication of feedback also appeared to be highly valued. For the section of students who preferred Feedback A, a strongest emphasis on content-related features was observed, especially conceptual information. However the `Depth' and `Evaluation' themes we rated lower than those who preferred the other types. Students who preferred Feedback B emphasized detailed discussion of the task content combined with critical evaluation of their solutions. In contrast, students who preferred Feedback C highlighted features associated with critical examination of their responses (strengths, weaknesses, and actionable suggestions) and the structured presentation of this feedback. Below, we detail these trends across the dataset as a whole and within each feedback type.

Across 1160 responses, students most frequently valued features associated with the Evaluation theme (69.4\%), followed by Content (55.2\%), Presentation (47.8\%), and Depth (44.2\%). Within Evaluation, Critique and Guidance were the most valued features. Within Content, Conceptual and Schema were rated most highly. For Presentation, Conciseness was emphasized most, while for Depth, the Detail feature was most frequently valued. Notably, no single feature was consistently valued across all three feedback types.

Students who ranked feedback from self-crafted prompts (Feedback A) most frequently emphasized conceptual arguments (the Conceptual feature) in their justifications. Specifically, 45.4\% of responses in this group evidenced the Conceptual feature, compared to 33.9\% for Feedback B and 26.5\% for Feedback C. This emphasis on conceptual reasoning aligns with the broader pattern observed in Fig.~\ref{fig:themes-combined}, where students who ranked Feedback A as most useful also valued the Content theme more than any other group (60.8\%). A similar, though less pronounced, trend was observed for the Clarity feature, with relatively low percentages across all groups (16.5\% for A, 7.6\% for B, and 17.5\% for C).

Students who preferred feedback generated from prompts incorporating features of prompt engineering (Feedback B) predominantly valued the `Content', `Depth', and to a slightly lesser extent the `Evaluation' themes compared to the other groups. In particular, students who chose Feedback B highlighted features associated with Affirmation (31\%), Correctness (17\%), Schema (39.5\%), and Detail (40.9\%) more than the other groups. They were also a close second in emphasizing Critique (46.8\%) and Guidance (47.4\%).

Students who preferred feedback generated from prompts combining foundational prompt-engineering techniques with principles of effective feedback (Feedback C) most often emphasized the `Evaluation' and `Presentation' themes. This is consistent with the design of the associated prompt, which explicitly asked for highlighting of strengths, shortcomings, and addressing gaps (Evaluation) as well as specifying a word limit and using a bulleted list (Presentation). These students also valued the `Depth' theme, though they were a close second to those who ranked Feedback B as their top choice. In contrast, they placed the least emphasis on the Content' theme compared to the other two groups.

Within this group, students most frequently highlighted Critique (48.4\%), Guidance (52.4\%), Conciseness (36.2\%), and Specificity (27.7\%). Smaller proportions also noted Clarity (17.5\%), Structure (18.2\%), and Intelligibility (18.9\%). Overall, these features were valued more highly by students in this group than by those in the other two groups.

\section{Discussion}\label{sec:discussion}
In the previous section, we presented results on the extent of sophistication evidenced in students’ prompts to AI when seeking feedback on their arguments about a physics task ({\bf RQ1}). We also discussed the feedback features and associated themes that emerged in students’ justifications of their preferred feedback choices among three options ({\bf RQ2}). The three feedback versions were generated using (i) self-crafted prompts (Feedback A), (ii) provided prompts incorporating foundational prompt-engineering techniques (Feedback B), and (iii) provided prompts combining foundational techniques with principles of effective feedback (Feedback C). Finally, we examined trends in students’ rankings of the three feedback types and the distribution of identified feedback features across them ({\bf RQ3}). The results across the research questions can be summarized as follows:

\begin{enumerate}
    \item Student-generated prompts rarely incorporate foundational prompting strategies in combination, as outlined in the prompt engineering literature.

    \item The emerging large language models frequently demonstrate ‘contextual intuitiveness’, i.e., taking into account the missing information such as the task’s context from prompts. Despite this sophistication, the models may not always provide conceptually accurate feedback.

    \item We identified four themes capturing the features students attribute for the perceived usefulness of AI-feedback: Evaluation, Content, Presentation, and Depth. The specific features include:: (i) Affirmation, (ii) Critique, (iii) Guidance, (iv) Correctness, (v) Conceptual, (vi) Schema, (vii) Clarity, (viii) Structure, (ix) Conciseness,  (x) Intelligibility, (xi)  Detail, and (xii) Specificity. 


    \item Those who preferred feedback generated from self-crafted prompts preferred conceptual-related information in feedback the most. Students who preferred feedback generated from prompt entailing prompt engineering techniques valued detailed discussion of conceptual-related information along with critical evaluation of their solutions. Students who preferred feedback generated from combining of prompt engineering techniques and principles of effective feedback valued the features of critical examination of their responses,  actionable suggestions along with their communication in a structured way.  There was no single feature that was consistently valued by students across the different preferred feedback types.  

    \item An overwhelming fraction of students preferred feedback generated from structured prompts, particularly the one entailing the combination of prompt engineering techniques and the principles of effective feedback. Students least preferred the feedback generated using self-crafted prompts. For a given preferred feedback choice, there was no common second preferred candidate.     
\end{enumerate}

The above results highlight several overarching aspects pertinent to the use of AI in STEM education. Firstly, students seldom employed best practices together when prompting AI. Despite this lack of sophistication, evolving AI models demonstrated notable `contextual intuitiveness' by rightly interpreting prompts based on overall content, even with ambiguous contexts. However, the need for well-structured prompts becomes evident in cases where AI responses to imperfect prompts were not well-positioned to facilitate students' learning. While techniques such as the use of delimiters may be less critical given the sophistication of current AI models, our findings underscore the importance of helping students craft well-structured prompts to maximize the potential of AI for their own learning.

Secondly, students overwhelmingly preferred feedback generated from structured prompts, particularly those embedding principles of effective feedback. Conversely, they showed the least preference for feedback generated from self-crafted prompts. This reinforces not only the importance of prompt engineering but also the value of leveraging education research on effective feedback in designing and delivering AI-driven feedback systems.

Thirdly, students identified a spectrum of valued features in AI feedback, ranging from conceptual discussions to structured communication. Yet, no single feature was consistently valued across all feedback types. Instead, each preferred type reflected unique combinations: students favoring Feedback A valued content-focused features; those preferring B valued content in combination with evaluative features such as strengths, shortcomings, and actionable suggestions; while those preferring Feedback C emphasized evaluative features along with structured presentation. These observations suggest a complex preference structure among students which warrants further investigation.  

These findings both complement and extend prior research on students’ use of AI in STEM contexts. While earlier studies have examined prompting strategies in domains such as embedded systems~\cite{sawalha2024analyzing}, programming~\cite{zhang2024students}, and chemistry problem solving by pre-service teachers~\cite{tassoti2024assessment}, our study extends this work into the domain of physics argumentation tasks. Sawalah {\em et al.}~\cite{sawalha2024analyzing} found that undergraduates primarily relied on three prompting strategies: directly copying-and-pasting questions, presenting a partially or completely phrased one-time question, and posing multiple questions. They also observed that 41\% of participants (23 of 56) used role-based prompting, particularly when posing multiple questions. Somewhat along with same lines, Tassoti~\cite{tassoti2024assessment} reported a predominant reliance of their participants on the `copy-and-paste’ approach when interacting with AI. Zhang {\em et al.}~\cite{zhang2024students} observed programming students preferring feedback derived from prompts that included their code, with feedback features such as specificity, clarity, and corrective nature guiding their preferences. 

Observations made in the current study align as well as differ from the above observations in several ways. Firstly, our task’s context involved students to make arguments about the physical significance of a graph and then seek feedback from AI. It naturally made the `copy-and-paste' approach an inefficient way of prompting. As a result, our data did not reflect this practice to the extent observed in the above studies, suggesting a possible relationship between task objectives and prompting approaches. Secondly, unlike the results from Sawalah et al., only 6 out of 1235 students in our data employed role-based prompting. However, consistent with Zhang et al., our participants preferred feedback from structured prompts that specified their arguments, attributing its usefulness to features such as Specificity, Clarity, Critique, and Guidance.

Valeri {\em et al.}~\cite{valeri2025exploring} similarly examined upper-secondary students’ prompting approaches to ChatGPT. They reported that 61\% of participants employed prompting strategies such as specifying output structure (e.g., bulleted lists), requesting detail or concise responses though explicit descriptors (e.g., `brief’, `longer’, or `medium’), and providing specific contexts to optimize results. Our observations show partial overlap with these findings. While two-thirds of our students specified context, we  observe requests for output format or explicit specification of detail levels. Interestingly, these aspects emerged in the valued feedback features students highlighted in their justifications.   

On feedback features, our findings also resonate with contemporary research. Henderson {\em et al.}~\cite{henderson2025comparing} qualitatively analyzed university students’ use of AI for soliciting feedback and identified valued features such as depth (nuanced vs. superficial), understandability, and contextualization (sensitivity to assignment requirements). A subset of our features—such as Specificity, Intelligibility, and Schema—align closely with these observations. Perhaps, the study from Prompiengchai {\em et. al}~\cite{prompiengchai2025practical} stands as the closest counterpart to our study in terms of integrating the principles of effective feedback into feedback from AI systems. The authors discuss the potential of leveraging prompting techniques such as Zero-shot and One-shot approaches in integrating the principles of effective feedback for students' formative assessment responses. Our study complements their discussion by not only blending prompting techniques with feedback principles, but by also capturing students' perspectives about its efficacy on feedback quality.     

\section{Conclusions, Limitations and Future Work}\label{sec:conclusions}

There is a growing body of research on leveraging AI platforms to facilitate students’ customized learning in physics, particularly through real-time feedback. This study contributes to these efforts by unpacking the spectrum of features students value in AI-generated feedback and their associated themes. We explore students’ preferences across three types of feedback generated from distinct prompt types: self-crafted prompts, provided prompts incorporating foundational prompt-engineering features, and provided prompts combining prompt-engineering principles with those of effective feedback. Finally, we also analyze students’ AI interactions by examining their prompts through the lens of foundational prompt-engineering techniques. Finally,  Detailed results are presented and discussed in Section~\ref{sec:discussion}.

Each result presented in this study offers a unique contribution to contemporary research on the use of AI systems for physics teaching and learning. As noted in Section~\ref{sec:intro}, while numerous studies have examined the efficacy of AI models in solving physics assessments and grading student responses, research on AI-generated feedback remains comparatively sparse. Among the few studies that focus on feedback, most employ empirical approaches that provide AI-driven feedback with little to no grounding in the theoretical principles of effective feedback from the education research literature. Noting this trend, there have been calls for blending theory into practice in the context of design and delivery of feedback in education settings. In addition to incorporating the principles, the current study also highlights students' clear preferences for feedback generated from prompts embedding these principles.

While prior research has considered students’ perceptions of linguistic quality or scientific accuracy in AI outputs~\cite{dahlkemper2023physics}, few studies have examined the features students associate with the perceived usefulness of feedback in the context of physics problem solving.  This study contributes to this literature by identifying twelve distinct features that students value in AI-generated feedback and grouping them into broad themes. Lastly, our study contributes to the literature by analyzing students’ prompts in terms of their sophistication through the lens of foundational prompting techniques.

The results reported in this study carry several implications for researchers. Firstly, students-attributed valued feedback features varied depending on their preferred feedback type, suggesting a potential complex structure guiding their choices. These results suggest the need for investigation focused on factors such as task's context, perceived task's difficulty, familiarity with AI models, or even attitudes toward AI use in learning environments on students' choices. We also observed students’ prompting strategies often fall short of expectations from the prompt-engineering literature. Future research could examine how students’ diligence in crafting prompts depends on factors such as task type (well-structured vs ill-structured) or participation format (individual vs. collaborative). Furthermore, while AI models demonstrated `contextual intuitiveness’, equity concerns remain regarding the educational divide concerning access to sophisticated models.

For instructors, our finding that students preferred feedback generated from structured prompts underscores the importance of blending AI literacy with content knowledge in physics learning environments. Such efforts can better prepare students to effectively integrate emerging technologies like AI into their learning. Our results also highlight the need to explicitly embed principles of effective feedback when leveraging AI models for designing and delivering feedback.

This study has several limitations. First, by design, students were aware of the prompts used to generate the three types of feedback. This may have introduced a bias in their preferences with notions such as - more sophisticated prompts yield more sophisticated feedback. Second, the study involved students’ interactions with AI on a single physics task. Future research using a wider range of tasks -- varying in type (multiple-choice, open-ended, conceptual, numerical, etc.) and quantity -- could yield deeper insights into students’ prompting approaches and feedback preferences. Third, students were restricted to crafting a single prompt for seeking feedback. In practice, AI interactions often involve iterative exchanges to refine responses, which was not captured in this study.

Future work will explore addressing the above limitations along with unpacking the co-occurrence of feedback features across preferred feedback types. Examining student preferences in relation to the correctness of their solutions also remains an avenue. Finally, we aim to investigate students’ perceptions of the usefulness of integrating AI-related activities into their assignments for their learning experiences.

\section{Acknowledgments}
ChatGPT-4o was used to correct grammar. This study was supported by Purdue University's Innovation Hub and U.S. National Science Foundation with grant number: 2111138.

%

\end{document}